\documentclass[pra,preprint,altaffilletter]{revtex4}
\usepackage{latexsym}
\usepackage{amsfonts}
\usepackage{amssymb}
\usepackage{amsmath}
\usepackage{graphicx}

\begin{document}

\title{Microwave spectroscopy on magnetization reversal dynamics of nanomagnets with electronic detection}
\author{J. Grollier}
\thanks{Now at Unit\'{e} Mixte de Physique CNRS/Thales, Route d\'{e}partementale 128, 91767 Palaiseau Cedex, France.}
\author{M. V. Costache}
\author{C. H. van der Wal}
\author{B. J. van Wees}
\affiliation{Physics of Nanodevices, Materials Science Centre, University of Groningen, Nijenborgh 4, 9747 AG Groningen, The Netherlands}
\date{\today }

\begin{abstract}
We demonstrate a detection method for microwave spectroscopy on magnetization reversal dynamics of nanomagnets. Measurement of the nanomagnet anisotropic
magnetoresistance was used for probing how magnetization reversal is resonantly enhanced by microwave magnetic fields. We used Co strips of 2 $\mu$m
$\times$ 130 nm $\times$ 40 nm, and microwave fields were applied via an on-chip coplanar wave guide. The method was applied for demonstrating
single domain-wall resonance, and studying the role of resonant domain-wall dynamics in magnetization reversal.
\end{abstract}

\maketitle

\section*{INTRODUCTION}

It is crucial for the implementation and miniaturization of magnetic and spintronic devices to understand the magnetization dynamics of
nanostructures at GHz frequencies. Our goal is to create and detect large amplitude ferromagnetic resonance \cite{fmr} (FMR) of individual
nanomagnets. This is of interest for realizing fast magnetization reversal, and for driving spin currents into adjacent normal metals
\cite{Brataas}. Cavity-based microwave techniques have been used for studying FMR, but these are not sensitive enough for studies of individual
nanomagnets and the dynamics of individual domain walls.
Gui \textit{et al.} \cite{Gui}, however, recently showed with a ferromagnetic grating that DC transport measurements on the ferromagnet can form a very sensitive probe for microwave induced FMR, charge dissipation, and their interplay.
Earlier experiments already showed that transport measurements also allow for probing the magnetic configuration of individual
submicron structures. Ono \textit{et al.} \cite{Ono} using the giant magnetoresistance (GMR) effect, and Klaui \textit{et al.} \cite{Klaui} using
the anisotropic magnetoresistance (AMR) effect, have detected domain wall motion in magnetic nanowires. Work on current-induced dynamics of a
single domain wall in a magnetic nanowire by Saitoh \textit{et al.} \cite{Saitoh} allowed for determining the domain wall mass. Further, the GMR
effect was used for real-time detection of the dynamics of spin valve devices \cite{Schumacher,Russek} and for observing spin-transfer induced
magnetic oscillations at GHz frequencies \cite{Kiselev}. We demonstrate here how the AMR effect can be used for detecting how microwave magnetic fields resonantly enhance magnetization reversal of individual nanomagnets that are embedded in electronic nanodevices. This allows for analyzing the magnetization dynamics in the metastable state prior to reversal of the magnetization.

\section*{EXPERIMENTAL REALIZATION}

We use devices that are patterned by electron beam lithography. In a first step, a gold coplanar waveguide (CPW) is defined with standard
lift-off techniques (Fig.\ref{fig1} (a)). The short at the end of the CPW forms a 2 $\mu$m wide microwave line, and provides the microwave
magnetic field. Then a device containing the nanomagnet is fabricated close to the microwave line with shadow mask techniques \cite{Jackel}. In
this paper we concentrate on the case of a cobalt strip of 2 $\mu$m $\times$ 130 nm $\times$ 40 nm. It is deposited by e-beam evaporation
parallel to the microwave line at 2 $\mu$m distance. In the same vacuum cycle, four aluminum fingers are deposited that form clean contacts with
the Co strip (Fig.\ref{fig1} (b)). The microwave field is perpendicular to the plane of the sample and the equilibrium direction of the
magnetization, which is a condition for driving the FMR \cite{Ebels}. The CPW is connected to a microwave signal generator via microwave probes
with 40 GHz bandwidth.

Our detection method of FMR is based on microwave-assisted magnetization reversal \cite{Thirion,Grundler}. Slowly sweeping a static magnetic
field parallel to the strip's long dimension is used for inducing a sudden switch event between the two saturated magnetic configurations. When
microwave-driven FMR occurs, the magnetic configuration is excited out of a metastable state, and the static-field induced switching occurs at
values closer to zero field. The switching fields are deduced from recording the strip's resistance $R(H)$ during the field sweep. When
approaching the switching field, the magnetization is pushed slightly out of its zero-field configuration, which causes a reduction of the
strip's AMR (the strip's AMR ratio is about 0.6 $\%$). Magnetization reversal is identified from a sudden return to the zero-field AMR value
(Fig.\ref{fig2}). The resistance of the sample is measured in a four probe geometry (see Fig.\ref{fig1}) with a lock-in detection technique and
5 $\mu$A AC bias current. All measurements are done at room temperature.

\section*{RESULTS AND DISCUSSION}

The switching of the samples is first characterized without applying a microwave field. In our particular sample, two types of $R(H)$ curves can
be obtained (Fig.\ref{fig2}). This can be understood when considering that in high-aspect-ratio samples as used here, magnetization reversal
occurs by domain wall nucleation and propagation \cite{McMichael,Ono}. The $R(H)$ curve $\bullet$ shows first a small reversible decrease of the
resistance \cite{reversible}, and then a sharp transition towards the initial resistance at $\approx$55 mT, noted as $up^{NoP}$. At this field a
domain wall propagates through the strip. For the $R(H)$ curve $\vartriangle$, the resistance also decreases progressively up to $dn^{P}$ at
$\approx$55 mT, but then drops sharply. $R$ is then constant up to $up^{P}$ at $\approx$65 mT, where a jump towards the initial value is
observed. In this case, instead of propagating directly through the sample, the domain wall gets pinned between the voltage probes (probably by
some defect arising from the lithographic process), and a higher field is needed to unpin the domain wall \cite{Klaui}. The decrease in
resistance $\Delta$R is due to the spin distribution in the domain wall, which gives a negative contribution to the AMR. By comparing $\Delta R$
to the total variation of resistance $\Delta R_{AMR}$, we can estimate the width of the domain wall by $W = d \Delta R / \Delta R_{AMR}\approx$
250 nm, with $d$=0.5 $\mu$m the distance between the voltage probes. This value is comparable to the width of domain walls observed in Co rings
of thickness and width similar to our sample \cite{Klaui2}.

We now turn to discussing microwave-assisted switching, measured in static field cycles while applying a microwave magnetic field as well. We
first set the amplitude of the microwave field  to a value of 2.2 mT \cite{fieldvalue}, and study the frequency dependence of the switching
fields. Fig.\ref{fig3}(a) shows results for $up^{NoP}$ and $dn^P$. The $up^{NoP}$ and $dn^P$ values are distributed over 0.5 mT due to thermal
broadening. In order to gain accuracy, the $R(H)$ curve for each frequency was performed 10 times and we plot the averaged values. Within the
precision of the measurement $up^{NoP}$ and $dn^P$ are equal: the value of the field at which the domain wall appears between the voltage probes
is the same for reversal with and without domain wall pinning. Further, we observe two resonances where the switching fields are decreased at
4.2 and 6.6 GHz. As in FMR measurements, the width and amplitude of these resonances are linked to the Gilbert damping parameter $\alpha$.

Fig.\ref{fig3}(b) shows how the switching fields $up^{NoP}$ and $dn^P$ depend on microwave amplitude $H_{MW}$, recorded for the frequencies 3,
4.2, and 6.6 GHz. The data taken at 3 GHz (outside the resonances in Fig.\ref{fig3}(a)) does not depend on $H_{MW}$. For the data at 4.2 and 6.6
GHz, however, the switching fields $up^{NoP}$ and $dn^P$ decrease linearly with $H_{MW}$. The precision of our measurement does not allow to
discriminate the 4.2 and 6.6 GHz curves. The same procedure is used to analyze the microwave dependence of $up^P$. Fig.\ref{fig3}(c) presents
results for $up^P$ vs. frequency. Here only one resonance is detected around 4.4 GHz. This behavior is confirmed in Fig.\ref{fig3}(d): The
switching field $up^P$ stays constant when $H_{MW}$ is increased for both 3 GHz and 6.6 GHz microwave fields. When the frequency of the
microwave field is set to 4.2 GHz, $up^P$ decreases with $H_{MW}$ with a step-like dependence.

We rule out that the observed phenomena are not FMR related but due to resonances in the microwave system. Resistance vs microwave amplitude at
high static magnetic field (200 mT), showed heating, but the frequency dependence at fixed amplitude showed variations less than 5 m$\Omega$.
With a microwave power of 14 dBm (corresponding to 2.2 mT) such resistance variations of the sample correspond to power variations in the
microwave line smaller than 1 dBm, and these cannot explain the large variations in switching fields that we observe (see the reference curves
at 3 GHz from Fig.\ref{fig3}(b) and (d) where the power is swept up to 18 dBm). We thus conclude that we observe FMR enhanced switching.

The interpretation of the results relies on the knowledge of the magnetic configuration before switching. At static fields slightly below $up^P$
the magnetic configuration is known: it consists of two domains separated by a pinned domain wall between the voltage probes. The magnetic
configuration at fields just inferior to $up^{NoP}$ and $dn^P$ is less clear: the magnetization in the sample can be close to uniform, or a
domain wall can already be nucleated, but outside of the voltage probes. Examination of the involved resonance frequency values shows that in
our experiments magnetization reversal is always initiated by domain wall dynamics, and not by the dynamics of the uniform mode. According to
the Kittel formula \cite{Kittel}, the resonance frequency of the uniform mode is : $f = \gamma_0 /
(2\pi)\sqrt{[H+(N_y-N_x)H_D][H+(N_z-N_x)H_D]}$. $N_{x,y,z}$ are the demagnetizing factors and $H_D$ the demagnetizing field. With
$N_{x,y}\approx t / w_{x,y}$, $N_z = 1 - N_x - N_y$, $H_D$ = 1.8 T and $H$ = - 60 mT, we find $f_{uniform}\approx$ 21 GHz. This is far from the
measured values, and the observed resonance frequencies also occur well outside the error margin for this estimate. The resonant mode for
$up^{NoP}$ and $dn^P$ at 4.2 Ghz is then more likely to be a domain wall resonance, just as for the 4.4 GHz resonance in $up^P$. To confirm this
last statement, we solve the following equations for domain wall motion \cite{Malozemoff}.

\begin{eqnarray}\label{DWmotion}
 \frac{\partial  \sigma }{\partial x}  =  \frac{M_s}{\gamma_0}
(\dot{\theta} + \alpha W^{-1} \dot{x}) &=& M_s H - M_s H_C
\frac{x}{x_c} \\
 \frac{\partial \sigma}{\partial \theta}  =  \frac{M_s}{\gamma_0} (-\dot{x} - \alpha W \dot{\theta})&=&  W H_D M_s
\sin\theta \cos\theta \nonumber \\
&-& W M_s H_{MW} \cos(\omega t)
\end{eqnarray}

Here $\sigma$ is the domain wall energy per unit area, $M_s$ the saturation magnetization, $\gamma_0$ the gyromagnetic ratio, $\omega$ the
microwave angular frequency, $x$ represents the domain wall displacement along the strip and $\theta$, the out-of-plane angle of the domain
wall, is a deformation parameter. The last term in Eq. (1) accounts for a quadratic pinning center of width $x_c$ and strength $H_C$
\cite{Baldwin}. For a constant domain wall width $W$ and small 
displacements, we calculate:

\begin{eqnarray}\label{results}
f &=& \frac{\gamma_0}{2\pi}\sqrt{\eta H_D H_C}\\
H_{SW} &=& H_C [1-\eta \frac{H_{MW}}{\alpha H_D}]
\end{eqnarray}

Here $f$ is the resonance frequency for the domain wall, with $\eta = W / x_c$. Using the values $H_D$ = 1.8 T, $H_C$ = 57.5 mT, and f = 4.2
GHz, we find with Eq.(3) that $\eta$ = 0.22. With $W$ = 250 nm this gives $x_c \approx$ 1 $\mu$m which is a reasonable value since the extension
of the potential well can be much larger than the physical dimensions of the pinning center \cite{Klaui}. Eq.(4) was obtained by using for the
switching condition the depinning of the domain wall at $x
> x_c$ and neglecting $H_C$ compared to $H_D$. This formula allows
us to fit the curve at 4.2 GHz of Fig.\ref{fig3}(b). Using the value $\eta$ = 0.22, the model fits the experimental data very well for $\alpha$
= 0.013, close to the 0.01 value measured in polycristalline cobalt \cite{Yuan}. As a conclusion, both the value of the resonance frequency (4.2
GHz) and the switching field dependence of $up^{NoP}$ and $dn^P$ on $H_{MW}$ at 4.2 GHz confirm that we see single domain wall resonance. We
also observed resonances around 4 GHz in smaller Co samples (600 nm $\times$ 130 nm $\times$ 20 nm) where the structure of the domain wall
should be similar to the one observed in 2 $\mu$m $\times$ 130 nm $\times$ 40 nm strips. When the domain wall is pinned between the voltage
probes, the dependence of the switching field $up^P$ is non-linear with respect to the amplitude of the microwave field. This can be explained
by strong oscillations in a non-quadratic pinning center. Additionally to the domain wall resonance at 4 GHz, we have observed a resonant mode
at 6.6 GHz. This resonance could be attributed to spin-waves or edges mode that can assist the onset of a reversal process.

\section*{CONCLUSIONS}

We have demonstrated a detection method for FMR in nanomagnets, based on transport measurements and microwave-assisted magnetization
reversal. We have used AMR measurements to probe how magnetization reversal of a Co strip is enhanced by resonant microwave magnetic fields. In this high-aspect ratio samples the magnetization reversal occurs by domain wall nucleation and propagation. This reversal mechanism is confirmed by our observations. Contrary to traditional FMR
techniques, the presented method allows to study single domain wall dynamics.

\section*{ACKNOWLEDGMENTS}

We acknowledge the RTN Spintronics Network, and the Stichting
Funtamenteel Onderzoek der Materie (FOM) for support.

\newpage
\begin{center}
FIGURE CAPTIONS\\
J.~Grollier \textit{et al.}
\end{center}

\begin{figure}[h]
\centering
\caption{(a) Optical microscope picture of the device including
the CPW with a short at the end.
(b) Scanning Electron Microscope picture of the Co strip, contacted by four Al
fingers.}
\label{fig1}
\end{figure}

\begin{figure}[h]
\centering
\caption{Resistance vs. static magnetic field $H$ curves measured
at room temperature. Here $H$ is parallel to the strip's longest dimension and slowly swept from -100 mT to +100 mT. With the same sample, two
behaviors $\bullet$ or $\vartriangle$ can be observed.}
\label{fig2}
\end{figure}

\begin{figure}[h]
\centering
\caption{(a) Average of $up^{NoP}$ ($\vartriangle$) and average
of $dn^P$ ($\bullet$) vs. frequency with a
2.2 mT microwave field. (b) Average of $up^{NoP}$ and $dn^P$  vs. H$_{MW}$ at
3 GHz $\blacksquare$, 4.2 GHz $\bullet$, 6.6 GHz $\vartriangle$. The line is the fit to the model.
(c) $\bullet$ : $up^{P}$ vs. frequency with a
2.2 mT microwave field (d) $up^{P}$  vs. $H_{MW}$ at
3 GHz $\blacksquare$, 4.2 GHz $\bullet$, 6.6 GHz
$\vartriangle$.}
\label{fig3}
\end{figure}


\begin{thebibliography}{99}
\bibitem{reversible}
This reversible decrease  can be due to a uniform rotation of
 the magnetization before the
  magnetic configuration breaks into
a domain wall, or to a domain wall pinned close to one of the voltage probes and inflating with the static magnetic field.
\bibitem{fieldvalue}
The effective value of the field was calculated considering the power sent into a perfect 50 Ohm load with a short a the end. Its value was also
checked by comparing DC and microwave heating of the sample (resistance readout) from sending power through the microwave line, and from
checking the microwave transmission of two inductively coupled CPW structures.
\end{thebibliography}
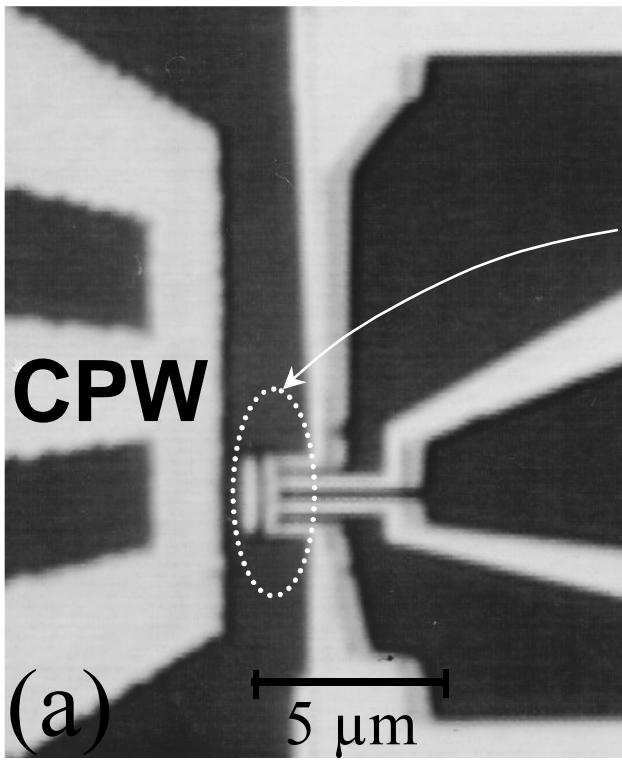
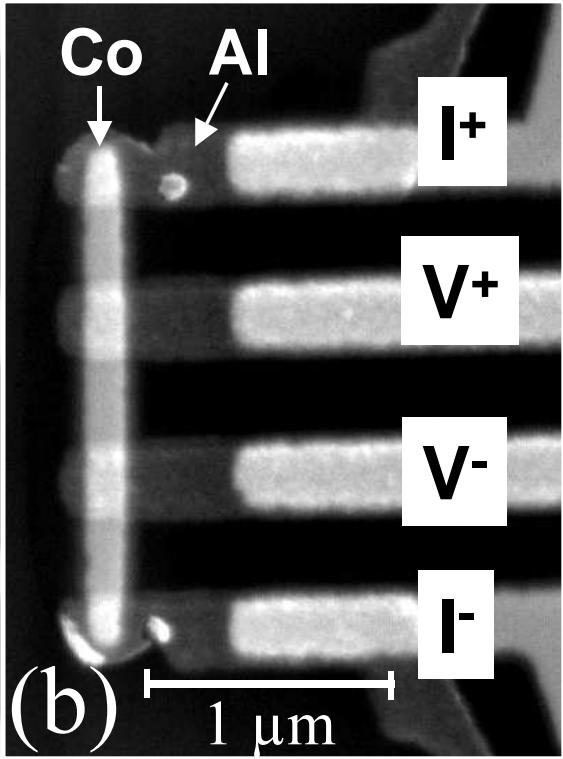
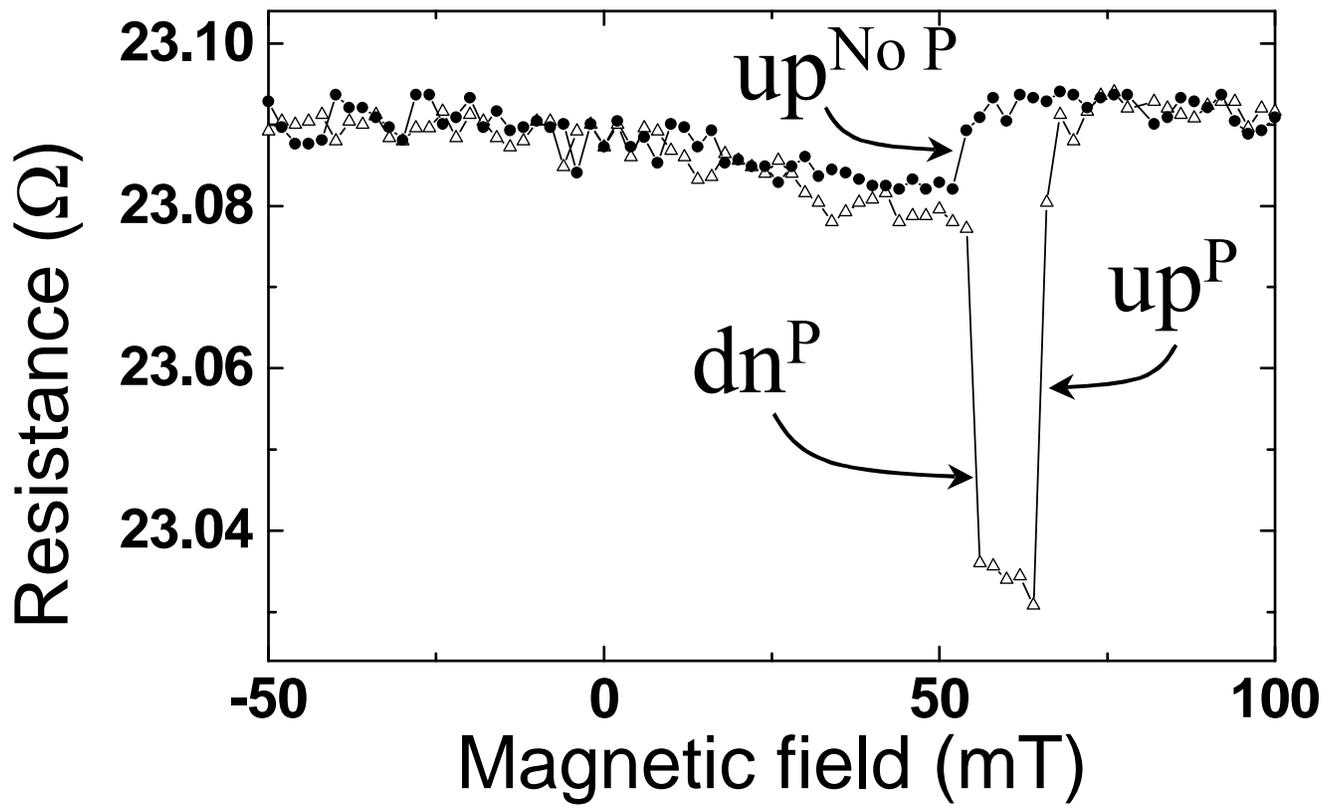
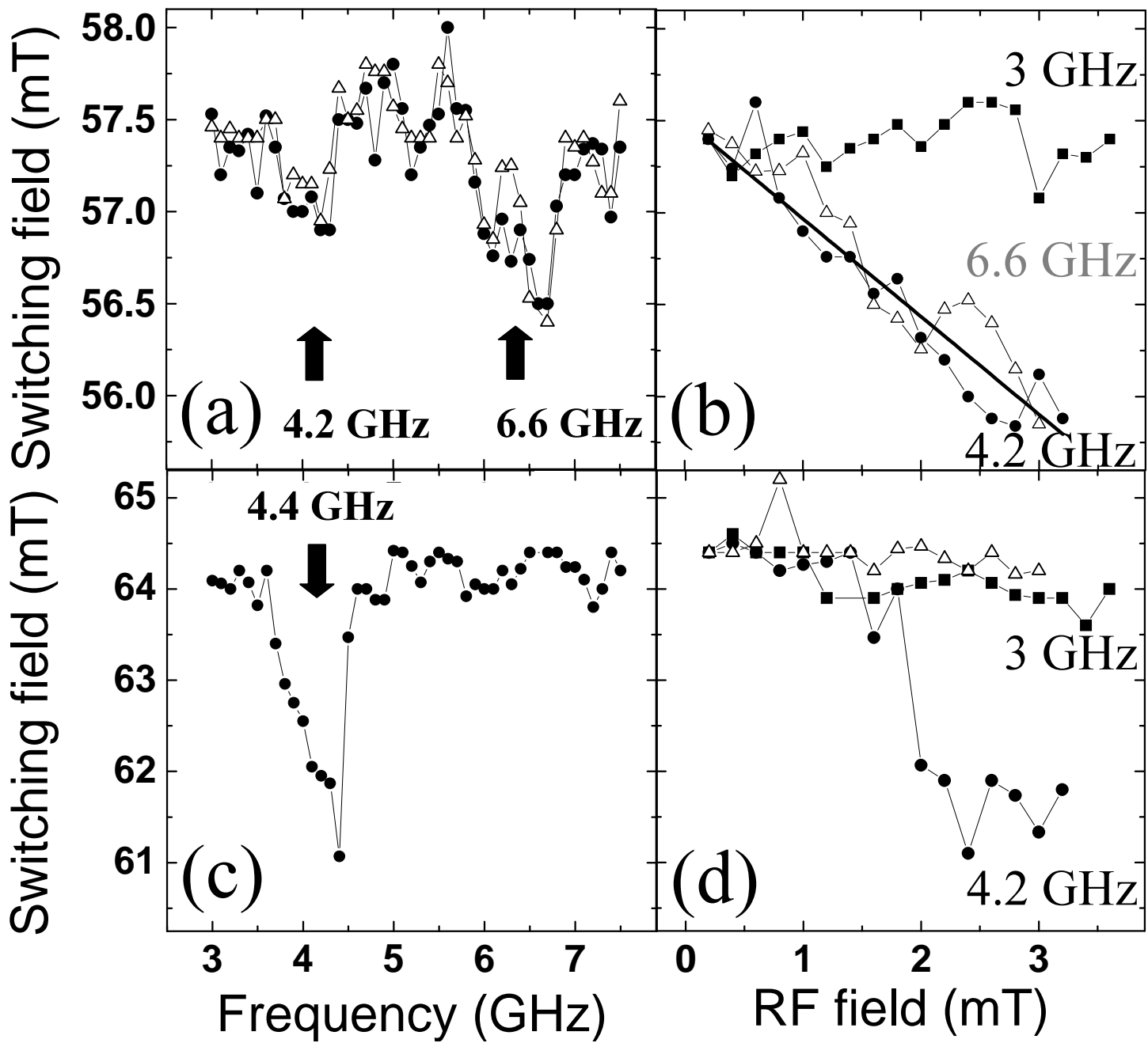
\end{document}